\documentclass[conference]{IEEEtran}
\IEEEoverridecommandlockouts 

\ifCLASSINFOpdf

\else

\fi

\usepackage{graphicx}

\usepackage{amssymb,amsfonts, amsmath, amsthm}
\usepackage{soul}
\usepackage{xcolor}
\usepackage{hyperref}

\usepackage{algorithm,algpseudocode}
\algrenewcommand\algorithmicindent{0.7em}
\usepackage{cite}
\usepackage[labelsep=space, font=small, labelfont=bf]{caption} 
\usepackage{subcaption}

\DeclareMathOperator*{\argmax}{arg\,max}

\begin{document}
%
\title{Integrating Lagrangian Neural Networks into the Dyna Framework for Reinforcement Learning
}


\author{
\IEEEauthorblockN{Shreya Das$^{*,1}$, Kundan Kumar$^{*,1}$, Muhammad Iqbal$^{*,1,4}$, Outi Savolainen$^2$, Dominik Baumann$^1$, \\ Laura Ruotsalainen$^2$, and Simo Särkkä$^{1,3}$}\thanks{$^{*}$Equal contribution.}
\IEEEauthorblockA{$^1$Department of Electrical Engineering and Automation, Aalto University, Finland} 
\IEEEauthorblockA{$^2$Department of Computer Science, University of Helsinki, Finland}
\IEEEauthorblockA{$^3$ELLIS Institute Finland}
\IEEEauthorblockA{$^4$VTT Technical
Research Centre of Finland}
\{shreya.das, kundan.kumar,  dominik.baumann, simo.sarkka\}@aalto.fi, muhammad.iqbal@vtt.fi \\
\{outi.savolainen, laura.ruotsalainen\}@helsinki.fi
}

\maketitle

\begin{abstract}
 Model-based reinforcement learning (MBRL) is sample-efficient but depends on the accuracy of the learned dynamics, which are often modeled using black-box methods that do not adhere to physical laws. Those methods tend to produce inaccurate predictions when presented with data that differ from the original training set. In this work, we employ Lagrangian neural networks (LNNs), which enforce an underlying Lagrangian structure to train the model within a Dyna-based MBRL framework. Furthermore, we train the LNN using stochastic gradient-based and state-estimation-based optimizers to learn the network's weights. The state-estimation-based method converges faster than the stochastic gradient-based method during neural network training. Simulation results are provided to illustrate the effectiveness of the proposed LNN-based Dyna framework for MBRL. 
\end{abstract}


\IEEEpeerreviewmaketitle

\section{Introduction}
\label{sec:intro}

Reinforcement learning (RL) employs a trial-and-error approach, letting an agent interact with the environment to learn to perform various tasks, such as solving decision-making problems or learning an optimal control policy \cite{sutton1998reinforcement}. In the last decade, RL has gained significant attention due to its successful applications across fields like robotics \cite{tang2025deep}, gaming \cite{li2025comprehensive}, control of multi-agent systems \cite{gao2020passivity}, and autonomous driving \cite{toromanoff2020end}. 

A widely used class of RL is model-free RL (MFRL), where the agent learns directly from interactions without requiring prior knowledge of system dynamics \cite{mnih2013playing, botvinick2019reinforcement}. 
 MFRL methods typically suffer from sample inefficiency \cite{haarnoja2018soft} and a slow convergence rate, which might not be problematic in simulated environments such as games, where interactions are inexpensive, but it severely limits applicability in real-world engineering systems such as robotics, 
autonomous vehicles \cite{wu2022uncertainty}, and quadrotors \cite{lambert2019low}. Data collection in these systems is costly and time-consuming, and mechanical systems are easily worn out due to extensive trials.

Model-based reinforcement learning (MBRL) improves sample efficiency by leveraging both real interaction data and a learned transition model \cite{sutton1996model}. There are various methods to solve the RL problem within the framework of MBRL \cite{nagabandi2018neural, hafner2019dream, ramesh2023physics, deisenroth2011pilco, feinberg2018model}. We use the Dyna framework, which alternates between real data collection and synthetic rollouts generated by a learned dynamics model \cite{sutton1991dyna, sutton1996model}.

Deep neural networks (DNNs) are commonly used to learn model of dynamical systems \cite{hafner2019dream}. However, using DNNs to learn the dynamics of systems with mechanical structure typically requires a large amount of interaction data, which runs counter to the objective of MBRL, where sample efficiency is critical \cite{pan2020trust}. A more suitable approach is to employ physics-informed models, such as 
Lagrangian neural networks (LNNs) \cite{cranmer2020lagrangian, ramesh2023physics}, which incorporate the underlying physical structure and therefore require significantly less data to identify the dynamics compared to generic black-box networks.
As we show in this paper, the training speed and overall performance of an LNN can be further improved by replacing the simple gradient-based optimization method with a state-estimation-based method \cite{haykin2004kalman}. A state-estimation-based optimization approach, such as one based on the extended Kalman filter (EKF), serves as a second-order neural network training method, where second-order information is evaluated recursively as a prior error covariance matrix \cite{haykin2004kalman, adeoye2024inexact}.
\begin{figure}
    \centering
\includegraphics[width=0.5\textwidth]{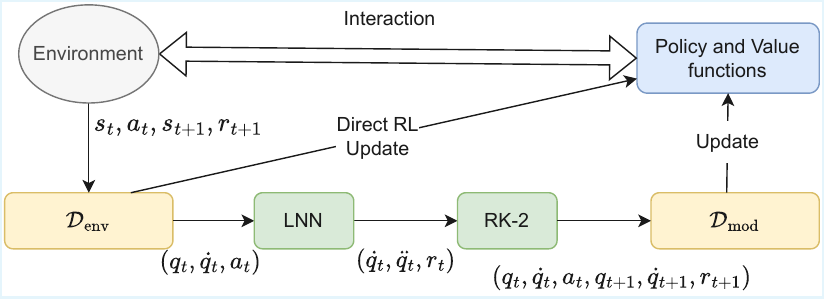}
    \caption{The basic idea behind the developed method of LNN-based MBRL in the Dyna framework. We learn the dynamics model using the LNN from real environment data and use an RK-2 integrator to generate model-based rollouts. The policy and value networks are trained on both real and model-generated data, improving sample efficiency while preserving physical consistency.}
    \label{fig:flow_diagram}
\end{figure}

The contributions of this paper are the following:
\begin{enumerate}
    \item We propose to utilize LNNs for MBRL, demonstrating increased sample efficiency than the state-of-the-art. 
    \item In addition, we employ state-estimation-based optimization methods for learning network weights, demonstrating even further improvements in efficiency.
    \item We also experimentally show that the proposed method outperforms the state-of-the-art physics-informed MBRL (PIMBRL) method in sample efficiency.
\end{enumerate}
The developed method is elaborated in Fig. \ref{fig:flow_diagram}.

\section{Integrating Lagrangian Neural Networks into the Dyna Framework}
In the following, we first state the RL problem and then outline the proposed Dyna-based LNN-MBRL approach.
Consider a Markov decision process (MDP) \cite{sutton1998reinforcement} defined by tuple $(\mathcal S, \mathcal A, p, r, \gamma, \rho_o)$, where $\mathcal S \subseteq \mathbb{R}^n$ and $\mathcal A \subseteq \mathbb{R}^m$ denotes the state and action space, $p(s_{t+1} \mid s_t, a_t)$ is the transition model, $r(s_t, a_t)$ is the reward function, $\gamma > 0$ is the discount factor, and $\rho_o(s_0)$ is the initial state distribution. The RL agent aims to solve the following optimization problem:
\begin{equation}
    \label{problem}
    \pi^* = \argmax_{\pi} \mathbb{E}_{\pi} \left\{\sum^{T_{max}}_{t= 0} \gamma^t r(s_t, a_t)\right\},
\end{equation}
where $\pi$ is any policy, $a_t \sim \pi(\cdot \mid s_t)$ is an action, and the next state is $s_{t+1} \sim p(\cdot \mid s_t, a_t)$.

To solve the problem given in \eqref{problem}, one can design an RL agent using policy-based, value-based, or model-based methods \cite{sutton1998reinforcement}. In this paper, we present a model-based method that improves sample efficiency. To that end, we develop an LNN to learn the dynamic model \cite{cranmer2020lagrangian} using less data than a DNN \cite{hafner2019dream} (see Section \ref{section_state_estimation}). To integrate model learning with RL methods which update the policy, one can use the Dyna framework \cite{sutton1991dyna}. The integrated LNN-based Dyna framework is provided in Section \ref{section_LNN_Dyna}. This framework integrates direct RL methods with model-based methods to update the policy or value function using synthetic data generated from the learned model, thereby improving sample efficiency. Using a DNN to learn dynamical systems requires a large dataset, which in the Dyna framework requires more interaction with the environment and thus results in poor sample efficiency. Next, we discuss the model-building part in the Dyna framework.


\section{State Estimation-Based Learning in LNNs}\label{section_state_estimation}
%
In this section, we present a state-estimation-based approach for learning the weights of the LNN. To do so, we first discuss the background of the LNN from the perspective of Lagrangian mechanics, and then move on to the weight learning.

\subsection{Lagrangian mechanics} 
In classical physics, the Lagrangian of a system can be modeled using coordinates, $(q,\dot{q})$, where $q(\tau)$ is the position variable and $\dot{q}(\tau)$ is the velocity variable at time $\tau$ \cite{goldstein19801}. 
The Lagrangian is expressed as
\begin{equation}\label{Lagrange_eqn}
  L(q,\dot{q}) = T(\dot{q}) - \Phi(q),
\end{equation}
where $\Phi(q)$ is the potential energy and $T(\dot{q})$ is the kinetic energy. Then, using the calculus of variations for minimizing the corresponding action integral, we get the Euler-Lagrange equations
\begin{equation}\label{EL_eqn}
    \frac{d}{d\tau} \frac{\partial L(q,\dot{q})}{\partial \dot{q}} - \frac{\partial L(q,\dot{q})}{\partial q} = a,
\end{equation}
where $a$ models the external forces. Using the chain rule in the above equation, we receive \cite{cranmer2020lagrangian}
\begin{equation}\label{qddot_formula}
    \ddot{q} = \left[ \frac{\partial^2 L}{\partial \dot{q} \partial \dot{q}} \right]^{-1} \, \left[ a+\frac{\partial L }{\partial q} - \frac{\partial^2 L}{\partial q \partial \dot{q}} \dot{q} \right]
    \triangleq \mathcal{G}[L].
\end{equation}
In our work, we employ the Lagrangian neural network approach, where a neural network with inputs $(q,\dot{q})$ and output $L$ is used. In particular, we use a neural network to approximate the Lagrangian $f_\omega(q,\dot{q}) \approx L(q,\dot{q})$, where $f_\omega(\cdot)$ is a neural network with learnable weights $\omega$ \cite{cranmer2020lagrangian}. Then, with the estimated $L$, we evaluate the acceleration variable, $\ddot{q}$, using the above equation. Thus, with the input of $(q,\dot{q})$, we get an estimate of $\ddot{q}$.  This is the overall structure of the LNN \cite{cranmer2020lagrangian}. 

Another approach would be to use Hamiltonian mechanics and the corresponding Hamiltonian neural networks (HNNs) \cite{greydanus2019hamiltonian}. However, HNNs would require replacing the state space with canonical coordinates $(q,p)$, where $p$ is the canonical momentum, which restricts the model's flexibility. Henceforth, in this paper, we use LNNs.

\subsection{Stochastic gradient-based optimization of LNN weights}
The basic idea of Lagrangian neural network training is to minimize a loss function $E$ defined as
\begin{equation}
\label{cost_function1}
    E(\omega) = \frac{1}{N} \sum_{i=1}^{N} \ell(y_i, \mathcal{G}[f_\omega](q_i,\dot{q}_i) ), 
\end{equation}
where $N$ is the size of the dataset, $y_i$ is the measurement of $\ddot{q}_i$, $\omega$ is the weight vector, $f_\omega(\cdot)$ is the neural network, $\mathcal{G}$ is the nonlinear operator defined in \eqref{qddot_formula}, and $\ell(\cdot)$ is a single-sample loss function, such as the mean square error \cite{bishop2006pattern, Goodfellow-et-al-2016}. The weights are updated stochastically using mini-batches of the dataset using the stochastic gradient-based method \cite{ruder2016overview, bottou2010large}. The general update rule for a mini-batch $\beta$ is
\begin{equation}\label{Eq_AdamWeight}
    \omega_{k+1} = \omega_k - \eta_k \nabla_{\omega } E_{\beta_k}(\omega_k), 
\end{equation}
where $\eta_k$ is the learning rate and $\nabla_{\omega } E_{\beta}$ is the gradient estimate from the mini-batch \cite{Goodfellow-et-al-2016}. A sufficiently small value of learning rate $\eta_k$ ensures the convergence of the method \cite{bottou2010large}, however, the convergence rate can be extremely slow.


\subsection{State estimation based optimization of LNN weights}
Unlike gradient-based optimization methods, where the weights of the neural networks are updated using deterministic gradient descent steps, state-estimation-based optimization methods treat the neural network weights as states of a dynamic system, and the training datasets are used as observations \cite{singhal1988training, haykin2004kalman, puskorius1991decoupled, adeoye2024inexact}. The neural network training problem now becomes a recursive Bayesian estimation problem that learns the posterior distribution of the weights given the observation datasets \cite{haykin2004kalman}.

We model the neural network weight dynamics using a random walk process to enable continual adaptation during model learning, allowing the weights to adjust as new observations become available \cite{haykin2004kalman,sarkka2023bayesian}. Specifically, the weight evolution is given by
\begin{equation}
    \omega_{i} = \omega_{i-1} + \mu_{i-1}, 
\end{equation}
where $\mu_{i-1} \sim \mathcal{N}(0, Q_{i-1})$ is the process noise. The observation can be modelled as 
\begin{equation}
    y_i = \mathcal{G}[f_{\omega_i}](q_i,\dot{q}_i) + \nu_i \triangleq h_i(\omega_i) + \nu_i,
\end{equation}
where $f_{\omega_i}(\cdot)$ is the neural network, and $\nu_i \sim \mathcal{N}(0, R_i)$ is the measurement noise. We can now utilize some well-established estimation approaches, such as the extended Kalman filter, cubature Kalman filter, or particle filter, to learn the weights $\omega_i$. In the EKF-based optimization \cite{haykin2004kalman} that we also use, the filter is applied to the data multiple times.

In this paper, we use the extended Kalman filter to learn the posterior distribution of weights \cite{haykin2004kalman, sarkka2023bayesian}. The weights are learned in two steps: the prediction step and the update step. In the prediction step, the marginal distribution of $\omega_i$ is $p(\omega_i \mid y_{1:i-1}) = \mathcal{N}(\omega_i \mid \hat{\omega}_{i\mid i-1}, \, P_{i\mid i-1})$, where  
\begin{align}
   \label{Eq:prior_mean} \hat{\omega}_{i|i-1}& =\hat{\omega}_{i-1|i-1},\\
   \label{Eq:prior_cov}   P_{i|i-1}& =P_{i-1|i-1}+Q_{i-1}.
\end{align}
Now, we use the measurement to evaluate the posterior distribution of $\omega_i$ given $y_{1:i}$, that is $p(\omega_i \mid y_{1:i}) \approx \mathcal{N}(\omega_i \mid \hat{\omega}_{i\mid i}, \, P_{i \mid i})$, where 
\begin{align}
   \label{Eq:Kalman_gain}  K_i & = P_{i\mid i-1}H_i^\top \left( H_i P_{i\mid i-1}H_i^\top+R \right)^{-1}, \\
   \label{Eq:posterior_mean}  \hat{\omega}_{i\mid i} &=\hat{\omega}_{i\mid i-1}+K_i \left(y_i - h_i(\hat{\omega}_{i\mid i-1}) \right), \\
  \label{Eq:posterior_cov}  P_{i\mid i} &= \left(I - K_i H_i \right) P_{i\mid i-1}.
\end{align}
Above, $H_i$ is the Jacobian matrix of $h_i$ computed at $\hat{\omega}_{i\mid i-1}$.
The state-estimation-based weight evaluation approach is advantageous when learning needs to be online, adaptive, and uncertainty-aware, especially under noisy or time-varying conditions.

\section{LNN-Based Dyna Framework for MBRL}\label{section_LNN_Dyna}
\begin{figure}
    \centering
\includegraphics[width=0.5\textwidth]{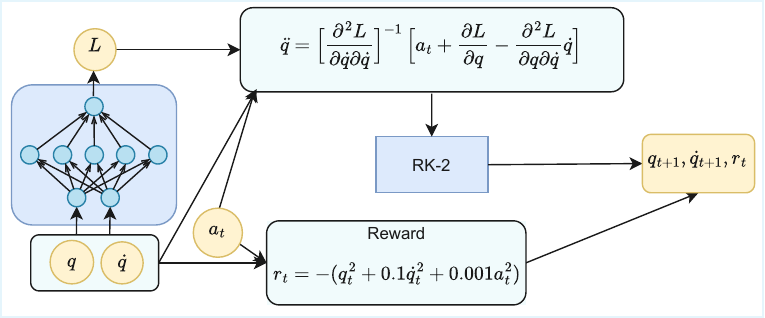}
    \caption{Model learning using LNN with $(q, \,\dot{q})$ as input and $L$ is the output of the network, which is further used to evaluate $\ddot{q}$ using the Euler-Lagrangian equation. The second-order Runge-Kutta (RK-2) method is then used to compute $(q_{t+1}, \, \dot{q}_{t+1})$. The reward is learning from the state-action pair.}
    \label{fig:lnn}
\end{figure}
In this section, we present the proposed LNN-based Dyna framework for MBRL, which represents a type of PIMBRL. In the Dyna framework, the agent interacts with the real environment, which yields $s_{t+1}$ and $ r_{t+1}$. The data obtained from the interactions with the real environment are stored in $\mathcal D_{\mathrm{env}}$. The reply buffer $\mathcal D_{\mathrm{env}}$ is used for two purposes: updating the policy and learning the system dynamics. In particular, we employ an LNN to model the dynamics. The incorporation of prior physical structure allows the LNN to learn the model in a data-efficient manner. The state vector, $s_t=(q_t,\dot{q}_t)$ consisting of the position $q$ and velocity $\dot{q}$ at time step $t$, and the action, $a_t$, are used as inputs to the LNN. Note that the Lagrangian depends only on the state $(q, \, \dot{q})$, and the angular acceleration, $\ddot{q}_t$, depends on the external generalized forces, $a_t$.
In this work, the network predicts the next step state, ($q_{t+1},\dot{q}_{t+1}$) and reward, $r_t$, from the present state ($q_{t},\dot{q}_{t}$) and the input torque, $a_t$. Recall that we can use the neural network to compute the estimate of the second derivative:
\begin{equation}\label{eq_GFw}
    \ddot{q} = \mathcal{G}[f_{\omega}](q, \, \dot{q}),
\end{equation}
where $\mathcal{G}$ depends on the generalized forces $a_t$.
The torque input $a_t$ acts as an external forcing term. In its absence, the system can only represent free dynamics, for instance, a pendulum swinging without any control input. With the torque input included, we can model controlled dynamics, which are essential for Dyna-based planning. 

Following \cite{ramesh2023physics}, we use the second-order Runge-Kutta (RK-2) method to get the states at the next time instant, $s_{t+1}=(q_{t+1},\dot{q}_{t+1})$ from $s_t = (\dot{q}_t,\ddot{q}_t)$, that is 
\begin{equation}\label{Eq_UpdateState}
    s_{t+1} = s_t + \Delta t \,(b_1 k_1 + b_2 k_2),
\end{equation}
where $\Delta t$ is the sampling time, $k_1$ and $k_2$ are computed as,
\begin{equation}\label{Eq_k1k2}
\begin{split}
    k_1= \begin{bmatrix}
        C \, s_t \\
        \mathcal{G}[f_{\omega}](s_t)
    \end{bmatrix}, \quad
    k_2= \begin{bmatrix}
      C \, (s_t+c\Delta tk_1) \\
    \mathcal{G}[f_{\omega}](s_t+c\Delta tk_1),
    \end{bmatrix},
\end{split}
\end{equation}
where $C = \begin{bmatrix} 0 & I \end{bmatrix}$ and 
\begin{equation*}
    c = \frac{2}{3}, \quad b_1 = \frac{1}{4}, \quad b_2 = \frac{3}{4}.
\end{equation*}
Please notice that it would also be possible to use other, higher-order Runge--Kutta methods \cite{Hairer+Norsett+Wanner:2008} instead, but we have adhered to the above method for consistency with \cite{ramesh2023physics}.
The learned model is then used to generate synthetic data, which is stored in $\mathcal D_{\mathrm{mod}}$. This is called model-based rollouts. The replay buffer, $\mathcal D_{\mathrm{mod}}$, is then used to update the policy or value function. This entire process is depicted in Fig. \ref{fig:flow_diagram} and Fig. \ref{fig:lnn}.

Inspired by \cite{liu2021physics}, we use an actor-critic-based RL agent. We assume that the policy and value function can be parametrized by $\pi(s_t; \theta_{\pi})$ and $V(s_t,a_t;\theta_{q})$, respectively. Here, $\theta_{\pi}$ and $\theta_{q}$ are weights of two neural networks. Samples from $\mathcal D_{\mathrm{env}}$ and $\mathcal D_{\mathrm{mod}}$ are used to update the policy network iteratively
\begin{equation}
    \label{policy_iteration}\theta_{\pi}^{k+1}=\theta_{\pi}^{k}+\alpha_{\pi}\nabla_{\theta_{\pi}}J(\theta_{\pi}^{k}),
\end{equation}
where $\alpha_{\pi}$ is the learning rate, and $\nabla_{\theta_{\pi}}J(\theta_{\pi})$ denotes the policy gradient with respect to the actor parameters $\theta_{\pi}$, obtained through the critic network
\begin{equation}
\label{ploicy_grad}
  \nabla_{\theta_{\pi}} J(\theta_{\pi}^{k})
= \mathbb{E}\!\left[\sum_{t=0}^{T} \nabla_{\theta_{\pi}} \log \pi(s_t;\theta_{\pi}^{k}) \cdot {V}(s_t,a_t;\theta_{q}^{k})\right].  
\end{equation}
The critic network $V(s,a;\theta_q)$ is updated by minimizing the temporal-difference loss function
\begin{equation}
    \theta_q^{*}=\arg\min_{\theta_q}\,\lVert V'_t-V(s_t,a_t;\theta_q)\rVert_{L_2},
\end{equation}
where $\lVert\cdot\rVert_{L_2}$ represents the $L_2$ norm and $V'_t$ is estimated based on the optimal Bellman equation
\begin{equation}
    V'_t=r_t+\gamma\,V \!\left(s_{t+1},\,{\pi}(s_{t+1};\theta_{\pi});\,\theta_q\right).
\end{equation}
The pseudo-code for the complete MBRL algorithm is presented in Algorithm \ref{Algo_Dyna}, the model-learning procedure is detailed in Algorithm \ref{Algo_Model}, and the EKF-based network weight update is provided in Algorithm \ref{Algo_Model_EKF}.

\begin{algorithm}[h!]
\caption{LNN-based Dyna framework for MBRL}
\label{Algo_Dyna}
\begin{algorithmic}[1]


\State Initialize randomly policy network $\pi$, value network $V$, transition model $\mathcal{G}[f_{\omega}]$, and replay buffers for both real $(\mathcal{D}_{\mathrm{env}})$ and simulated $(\mathcal{D}_{\mathrm{mod}})$ environments.
\State Initialize state $(s_0)$, reward $(r_0)$, and done signal $(d_0)$ randomly for real environment.

\For{episode $= 1, \ldots, M$}
    \For{$i = 0, \ldots, T$}
        \State Evaluate $a_i = \pi(s_i;\theta_\pi)$ in real environment.
        \State Store $(s_i, a_i, s_{i+1}, r_i, d_i)$ in replay buffer, $\mathcal{D}_{\mathrm{env}}$.
        \If{episode ends}
            \State Reset environment
        \EndIf
    \EndFor

    \If{$\text{size}(\mathcal{D}_{\mathrm{env}}) > \text{threshold}$}
        \State Sample batch of real state-action pair from $\mathcal{D}_{\mathrm{env}}$.
        \State Update the transition model ${\mathcal{G}}[f_{\omega}]$ (Algorithm \ref{Algo_Model}) using
        \Statex \quad \,  the data loss on batch $\mathcal{D}_{\mathrm{env}}$.
    \EndIf

    \If{data loss $< \text{threshold}$}
        \For{$j = 1, \ldots, L_M$}
            \State Sample a batch of $s_j$ from $\mathcal{D}_{\mathrm{env}}$ and $\mathcal{D}_{\mathrm{mod}}$.
            \State Apply actions $\{a_j = \pi(s_j;\theta_\pi)\}_{j=1}^{n_b}$ on transition \Statex \qquad   \,
            model ${\mathcal{G}}[f_{\omega}]$.
            \State Store $\{(s_j, a_j, s_{j+1}, r_{j+1}, d_{j+1})\}_{j=1}^{n_b}$ in $\mathcal{D}_{\mathrm{mod}}$.
        \EndFor
    \EndIf

    \If{$\text{size}(\mathcal{D}_{\mathrm{env}})$ $\&$
        $\text{size}(\mathcal{D}_{\mathrm{mod}}) > \text{threshold}$}
        \State Sample batch of state action pair from $\mathcal{D}_{\mathrm{mod}}$.
        \State Update model $\mathcal{G}[f_{\omega}]$ using physical loss on sampled 
        \Statex \quad \, state action pair.
    \EndIf

    \State Sample batch of state, action, reward, and done signal \Statex \quad 
    from $\mathcal{D}_{\mathrm{env}} \cup \mathcal{D}_{\mathrm{mod}}$.
    \State Update policy network $\pi$ and value network $V$ on the \Statex \quad 
    sampled state-action pair using Algorithm 3 in \cite{liu2021physics}.
\EndFor

\end{algorithmic}
\end{algorithm}

\begin{algorithm}[h!]
\caption{Algorithm for model learning}
\label{Algo_Model}
\begin{algorithmic}[1]
\State \textbf{Function:} 
$\big[q_{t+1}, \dot{q}_{t+1}, r_t\big] = \text{Model-Learn}\big(q_t, \dot{q}_t, a_t\big)$
\State Learn the LNN weights using \eqref{Eq_AdamWeight} or Algorithm~\ref{Algo_Model_EKF}. 
\State Compute $\ddot{q}_t$ using $\mathcal{G}[f_\omega]$.
\State Evaluate $k_1$ and $k_2$ from \eqref{Eq_k1k2}.
\State Update the state $s_{t+1} = (q_{t+1}, \dot{q}_{t+1})$ using \eqref{Eq_UpdateState}.
\State Compute reward $r_t = -\left(q_t^2 + 0.1\,\dot{q}_t^2 + 0.001\,a_t^2\right)$.
\State \Return updated state, $s_{t+1}$ and reward, $r_t$. 
\end{algorithmic}
\end{algorithm}

\begin{algorithm}[h!]
\caption{Algorithm for model learning using EKF}
\label{Algo_Model_EKF}
\begin{algorithmic}[1]
\State Initialize $\hat{\omega}_0$, $P_{0|0}$ and choose $Q$ and $R$.
\For{$i = 0, \ldots, T$}
    \State Predict prior mean $\hat{\omega}_{i|i-1}$ and covariance $P_{i|i-1}$ using
   \Statex \quad   \eqref{Eq:prior_mean} and \eqref{Eq:prior_cov}, respectively.
    \State Compute $\hat{y}_{i|i-1} = h_i(\hat{\omega}_{i|i-1})$.
    \State Compute Kalman gain $K_i$ using \eqref{Eq:Kalman_gain}.
    \State Update the posterior weight $\hat{\omega}_{i|i}$ and its covariance $P_{i|i}$
    \Statex \quad  using \eqref{Eq:posterior_mean} and \eqref{Eq:posterior_cov}, respectively. 
\EndFor
\end{algorithmic}
\end{algorithm}

\section{Simulation Results}

In this section, we evaluate the proposed LNN-based MBRL on an inverted pendulum problem \cite{liu2021physics} using the OpenAI Gym environment \cite{brockman2016openai}. The system state consists of the pendulum angle and angular velocity, while the model dynamics, control torque constraints, and associated parameters follow the formulation in \cite{liu2021physics}. The results can be extended for multi-degree-of-freedom articulated mechanical systems as the LNN-based dynamics model scales by increasing the dimensionality of the generalized coordinates.  The control objective is to stablize the pendulum around the upright position ($q \approx 0$) while minimizing control effort. The reward function is defined as 
\begin{equation}
    r_t = -(q_t^2 + 0.1 {\dot{q}_t}^2 + 0.001 {a_t}^2), 
\end{equation}
which assigns higher rewards near the upright equilibrium with small control inputs.

\begin{figure}
    \centering
    \includegraphics[width=7.5 cm, height=5cm]{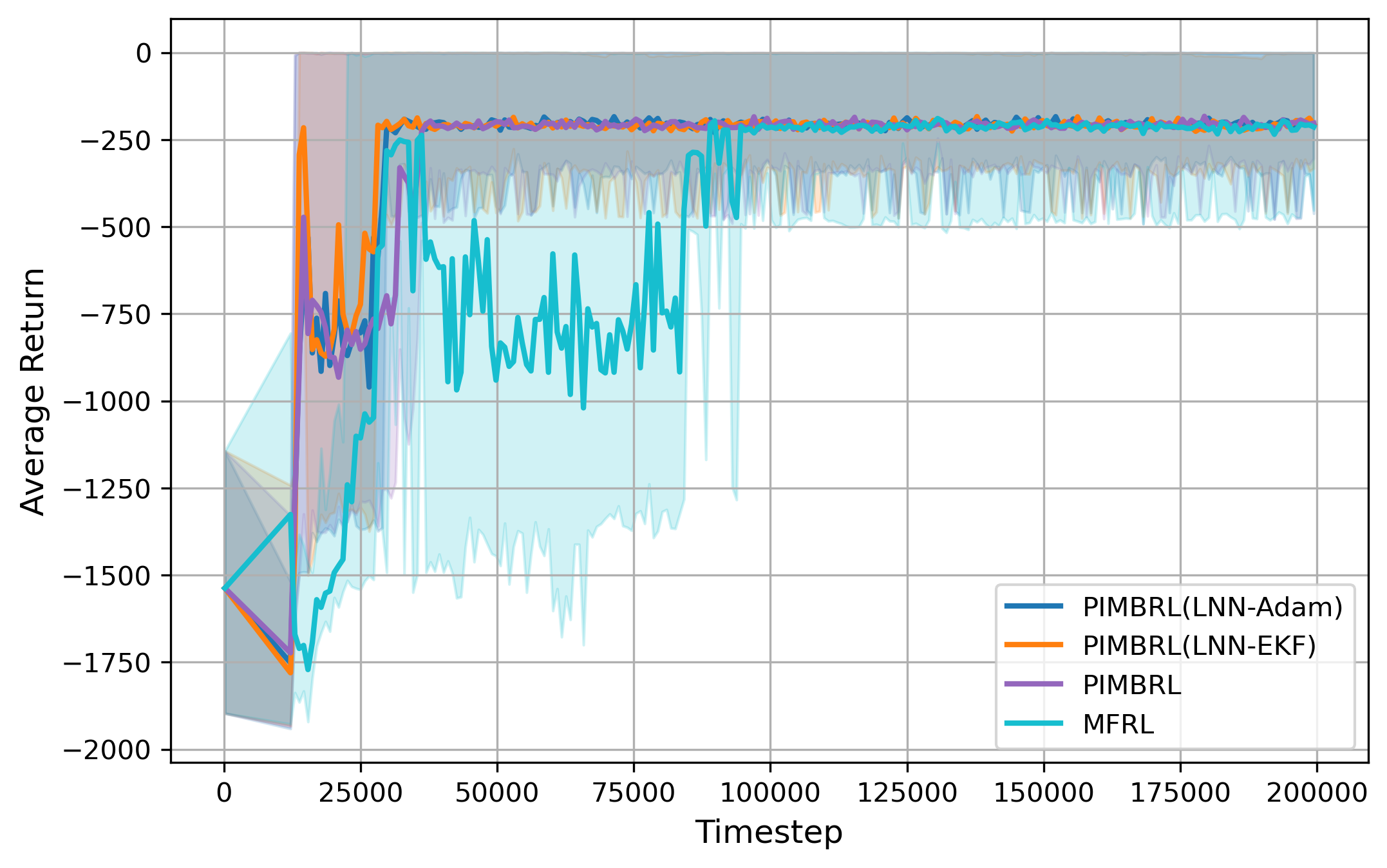}
    \caption{Average return versus timestep plot for the proposed PIMBRL using LNN with Adam and EKF as optimizers, PIMBRL using DNN with constraints as in \cite{liu2021physics}, and MFRL.}
    \label{fig:AvgReturn}
\end{figure}

We compare our proposed PIMBRL algorithm, which uses the LNN trained with Adam and EKF, with PIMBRL using a constrained DNN as in \cite{liu2021physics} and with a model-free RL (MFRL) baseline on the inverted pendulum problem.
The average returns are shown in Fig.~\ref{fig:AvgReturn}. 
From the figure, it can be observed that the average return of PIMBRL with LNN using Adam reaches approximately $-200$ after about 30,000 timesteps, while PIMBRL with LNN using EKF achieves the same performance slightly earlier, at about 28,500 timesteps. The PIMBRL variant with a constrained DNN, following \cite{liu2021physics}, reaches the target return of $-200$ near 36,500 timesteps. In contrast, the MFRL baseline fluctuates around $-750$ for a prolonged period and converges to $-200$ only after nearly 90,000 timesteps. 
These results demonstrate the superior sample efficiency of PIMBRL. The incorporation of physics-informed structure into the LNN and the use of state-estimation-based optimizers, such as the EKF, enable the model to learn the system dynamics more effectively with fewer interactions.

\section{Conclusion}
In this paper, we have combined a Lagrangian neural network with the Dyna framework. The resulting method enforces an underlying Lagrangian structure during neural network training in the Dyna framework of MBRL. The LNN used in the proposed method helps address finite-data-related challenges, which are generally encountered in black-box-based dynamic models, especially when operating at the edge of the training data. In this work, the LNN is trained using state-estimation-based optimization methods to learn its weights. The EKF optimizer leverages curvature information and adaptive scaling, leading to faster, more stable convergence than conventional first-order gradient-based optimizers.


\bibliographystyle{IEEEtran}
\bibliography{refs}

\end{document}